# Growth, Annealing Effects on Superconducting and Magnetic Properties and Anisotropy of FeSe$_{1-x}$Te$_x$ (0.5 ≤ x ≤ 1) Single Crystals

Takashi Noji, Takumi Suzuki, Haruki Abe, Tadashi Adachi, Masatsune Kato, and Yoji Koike

Department of Applied Physics, Tohoku University, Sendai 980-8579

Single crystals of FeSe$_{1-x}$Te$_x$ (0.5 ≤ x ≤ 1) have been grown by the Bridgman method. After annealing them at 400℃ for 100 h in vacuum, single crystals of x = 0.5 − 0.9 have exhibited bulk superconductivity. Anisotropic properties of the electrical resistivity and upper critical field, $H_{c2}$, have been investigated for the single-crystal FeSe$_{1-x}$Te$_x$ with x = 0.6. It has been found that the in-plane resistivity, $\rho_{ab}$, shows a metallic temperature-dependence, while the out-of-plane resistivity, $\rho_c$, shows a broad maximum around 100 K. The resistivity ratio, $\rho_{ab}/\rho_c$, is 44 and 70 at 290 K and just above the superconducting transition temperature, $T_c$, respectively. The anisotropic parameter, $\gamma \equiv H_{c2}^{\parallel}/H_{c2}^{\perp}$ (The superscripts $\parallel$ and $\perp$ indicate field directions parallel and perpendicular to the ab-plane, respectively.), is estimated as 2.7 just below $T_c$.

Ⅰ. Introduction

The discovery of superconductivity at 26 K in the iron-based oxypnictide superconductor LaFeAsO$_{1-x}$F$_x$ is quite surprising,[1] since compounds including iron are usually so magnetic as not to be superconducting. To the surprise, moreover, the superconducting transition temperature, $T_c$, has been raised up to 55 K by replacing La with other rare-earth elements such as Sm.[2] After that Hsu *et al*. have discovered superconductivity with $T_c$ = 8 K in the PbO-type structure FeSe.[3] This has also attracted great interest, because the crystal structure is so simple as to be composed of a stack of edge-sharing FeSe$_4$-tetrahedra layers which are similar to FeAs$_4$-tetrahedra layers in LaFeAsO$_{1-x}$F$_x$, and because $T_c$ of FeSe has markedly increased up to 37 K by the application of high pressure of 7 GPa.[4] It has been found that $T_c$ of FeSe increases through the partial substitution of Te for Se as well, shows a maximum 14 K at x = 0.6 in FeSe$_{1-x}$Te$_x$ and the superconductivity disappears at x = 1, namely, in FeTe.[5,6] The end member FeTe is not superconducting but develops an antiferromagnetic order at low temperatures below ~ 67 K where a tetragonal-to-monoclinic structural phase transition occurs.[7,8] As for the single-crystal growth of FeSe$_{1-x}$Te$_x$, tiny crystals of FeSe with a size of ~ 500 μm and $T_c$ = 10.4 K have been obtained using a NaCl/KCl flux.[9] Recently, large-sized single-crystals of Fe$_{1+y}$Te and Fe$_{1+y}$Se$_{1-x}$Te$_x$ with the tetragonal structure have successfully been grown by the Bridgman method.[10-12]

In this paper, we report on the growth and characterization of Fe$_{1+y}$Se$_{1-x}$Te$_x$ (0.5 ≤ x ≤ 1) single crystals with y values as small as possible due to excess Fe randomly occupying the so-called Fe(2) site in the crystal structure between neighboring square planar sheets of Fe.[7] Annealing effects on the superconducting and magnetic properties of Fe$_{1+y}$Se$_{1-x}$Te$_x$ (0.5 ≤ x ≤ 1) single crystals are also investigated. Moreover, anisotropic properties of the electrical resistivity and upper critical field, $H_{c2}$, of the annealed single-crystal FeSe$_{1-x}$Te$_x$ with x = 0.6 are investigated.

Ⅱ. Experimental

Single crystals of FeSe$_{1-x}$Te$_x$ were grown by the Bridgman method. Starting materials were powders of Fe (purity 3N), Se (purity 3N) and Te (purity 4N). The powders prescribed in the nominal composition described in Table Ⅰ were thoroughly mixed in an argon-filled glove box. The mixed powders were sealed in an evacuated quartz tube. Since the quartz tube often cracked upon cooling, the tube was sealed into another large-sized evacuated quartz tube. The doubly sealed quartz ampoule was stood in a furnace so that single crystals were grown using the temperature gradient in the furnace. The ampoule was heated at 600℃ for 100 h and successively at 950 - 1050℃ for 30 h, and then cooled down to 650℃ at the rate of 2 - 3℃/h, followed by furnace cooling down to room temperature.

Grown crystals were characterized by the x-ray back-Laue photography and the powder x-ray diffraction. The chemical composition was determined by the inductively coupled plasma atomic emission spectroscopy (ICP-AES). The composition of the surface of the crystals was checked using an electron probe microanalyzer (EPMA).

The magnetic susceptibility, $\chi$, was measured using a superconducting quantum interference device (SQUID) magnetometer (Quantum Design, Model MPMS). Measurements of the electrical resistivity were carried out by the standard DC four-probe method. The anisotropy in the resistive superconducting transition under magnetic field was measured using a commercial apparatus (Quantum Design, Model PPMS) to investigate the anisotropy of $H_{c2}$. Here, the magnetic field was always applied perpendicular to the direction of the excitation current flowing in the ab-plane.

## Ⅲ. Results and Discussion

We have succeeded in growing sizable single-crystals of $0.6 \leq x \leq 1$, but it was hard to obtain single crystals with $x = 0.5$ whose dimensions of the ab-plane are larger than 1 mm×1 mm. As-grown crystals of $0.6 \leq x \leq 1$ were easily cloven. Figures 1 (a) and (b) show a picture of as-grown single-crystals with cleavage surface and a x-ray back-Laue photograph in the x-ray perpendicular to the cleavage surface, respectively. The fourfold symmetry in the back-Laue photograph is due to the tetragonal structure, indicating that the c-axis is perpendicular to the cleavage surface. The powder x-ray diffraction has revealed that the obtained crystals are of the single phase without any impurity phases. Compositions of the obtained crystals chemically analyzed by ICP-AES are listed in Table Ⅰ. It is found that the chemical compositions of the single crystals are in approximate agreement with the nominal ones. Experimentally, it was hard to grow stoichiometric crystals of $FeSe_{1-x}Te_x$ without excess Fe.

Figure 2 shows the temperature dependence of the in-plane resistivity, $\rho_{ab}$, of as-grown single-crystals of $FeSe_{1-x}Te_x$ ($0.5 \leq x \leq 1$). It is found that superconductivity appears in these samples except for $x = 1$. Values of $T_c$ are comparable with those of sintered polycrystalline samples.[5,6] As for x = 1, namely FeTe, $\rho_{ab}$ shows a semiconductor-like behavior at high temperatures, whereas $\rho_{ab}$ drops steeply at ~ 65 K and then exhibits a metallic behavior at low temperatures. The discontinuous change in $\rho_{ab}$ is due to the structural phase transition accompanied by the magnetic transition.[7,8]

Figure 3 shows the temperature dependence of $\chi$ in a low magnetic field of 1 mT parallel to the c-axis for as-grown and annealed single-crystals of x = 0.5 − 0.9. It is found that as-grown crystals of x = 0.5 and 0.6 display diamagnetism due to bulk superconductivity, while as-grown crystals of x = 0.7 - 0.9 do not. That is, the resistive superconducting transition observed in as-grown crystals of x = 0.7 − 0.9 is not due to bulk superconductivity. The temperature dependence of $\chi$ for as-grown single-crystals of x = 0.8 − 1 in a high magnetic field of 1 T parallel to the c-axis is shown in Fig. 4. It is found that the magnetic transition temperature, $T_m$, defined at the temperature where $\chi$ rapidly changes, decreases with decreasing x. These results suggest that a phase separation into filamentary superconducting regions and non-superconducting regions takes place in as-grown single-crystals of x = 0.7 - 0.9, which is consistent with the preceding result by Sales et al.[11]

We have annealed the as-grown crystals at 400℃ for 100 h in vacuum. As seen in Fig. 3, annealed crystals of x = 0.5 − 0.9 exhibit bulk superconductivity, whereas as-grown crystals of x = 0.7 − 0.9 do not. Even for x =0.5 and 0.6, the diamagnetic signal is enhanced through the annealing. Values of $T_c$, defined as onset temperature of the Meissner effect, are estimated as 14.0 K, 14.2 K, 14.3 K, 13.8 K and 11.3 K for annealed crystals of x = 0.5, 0.6, 0.7, 0.8 and 0.9, respectively. As seen in Fig.4, $T_m$ of x = 1 increases a little up to 69 K through the annealing, while the magnetic transition of x = 0.9 disappears. Dependences on x of $T_c$ and $T_m$ for as-grown and annealed crystals are shown in Fig. 5. It seems that the distribution of Se and Te in a crystal becomes homogeneous through the annealing as pointed by Taen et al.,[12] so that the antiferromagnetic order observed for as-grown crystals of x = 0.8 and 0.9 disappears and alternatively bulk superconductivity appears for annealed single-crystals of x = 0.7 − 0.9. However, no structural change through the annealing could be detected by the powder x-ray diffraction. In addition, no change of the composition of the surface of the crystals through the annealing was detected by the EPMA measurements.

Figure 6 shows the temperature dependence of $\rho_{ab}$ and the out-of-plane resistivity, $\rho_c$, and the resistivity ratio, $\rho_c/\rho_{ab}$, for the single crystal of x = 0.6 annealed at 400℃ for 100 h in vacuum.

The $\rho_{ab}$ shows a metallic temperature-dependence at low temperatures below 140 K, while it is nearly constant at high temperatures above 140 K. This behavior of $\rho_{ab}$ is almost the same as that observed by Taen et al.[12] in the single-crystal FeTe$_{0.61}$Se$_{0.39}$. On the other hand, $\rho_c$ is semiconductor-like at high temperatures, shows a broad maximum around 100 K and changes to be metallic at low temperatures below ~ 100 K. The resistivity ratio increases with decreasing temperature, while it is nearly constant at low temperatures below ~ 50 K. The resistivity ratio is 44 and 70 at 290 K and just above $T_c$, respectively. These values are much smaller than those of the high-$T_c$ cuprate Bi$_2$Sr$_2$CaCu$_2$O$_8$ whose single-crystals are cloven as easily as FeSe$_{1-x}$Te$_x$ single crystals.[13] The anisotropic behaviors of $\rho_{ab}$ and $\rho_c$ are similar to those observed in the layered perovskite Sr$_2$RuO$_4$.[14,15] As in the case of Sr$_2$RuO$_4$, this compound is regarded as a two-dimensional metal at high temperatures above ~ 100 K, while it becomes an anisotropic three-dimensional metal at low temperatures below ~ 100 K.

Figure 7 shows the angular dependence of $\rho_{ab}$ in various constant fields at a temperature just below $T_c$ for the single crystal of x = 0.6 annealed at 400°C for 100 h in vacuum. The $H_{c2}$ is defined at the intersection between $\rho_{ab}$ and the half of the normal-state value of $\rho_{ab}$ in zero field. The angular dependence of $H_{c2}$ obtained thus is shown in Fig. 8. This is well expressed as $H_{c2}(\theta) = H_{c2}^{//}(\cos^2\theta + \gamma^2\sin^2\theta)^{-1/2}$, based on the effective mass model.[16] Here, $\theta$ is the angle between the ab-plane and the field direction and $\gamma$ is the anisotropy parameter defined as $\gamma \equiv H_{c2}^{//}/H_{c2}^{\perp}$. Here, the superscripts $//$ and $\perp$ indicate the field direction parallel and perpendicular to the ab-plane, respectively. The fitting parameter $\gamma$ obtained is 2.7.

This value of $\gamma$ is larger than those of FeSe$_{1-x}$Te$_x$ reported so far. That is, $\gamma$ is estimated as 1.6 for FeSe$_{1-x}$Te$_x$ with x = 0.7 by Chen et al.[10] and as less than 2 near $T_c$ and decreasing with a decrease of temperature for Fe$_{1.11}$Se$_{1-x}$Te$_x$ with x = 0.6 by Fang et al.[17] Our large value of $\gamma$ may be due to the good quality of our crystal owing to the annealing. In any case, our value of $\gamma$ is similar to those of Ba$_{1-x}$K$_x$Fe$_2$As$_2$[18,19] and much smaller than the smallest value of $\gamma$ = 5 in YBa$_2$Cu$_3$O$_y$ among the high-$T_c$ cuprates.[20] This suggests that the Fermi-surface topology of FeSe$_{1-x}$Te$_x$ is similar to that of Ba$_{1-x}$K$_x$Fe$_2$As$_2$ and more or less three-dimensional and different from that of

high-$T_c$ cuprates.

Finally, it is noted that the value of $\gamma$ is able to be estimated from the value of $\rho_c/\rho_{ab}$ as $\gamma \equiv H_{c2}^{\parallel}/H_{c2}^{\perp} = (m^*_c/m^*_{ab})^{1/2} = (\rho_c/\rho_{ab})^{1/2}$, assuming that $\rho_{ab}$ and $\rho_c$ are given by $m^*_{ab}/ne^2\tau$ and $m^*_c/ne^2\tau$, respectively. Here, n is the carrier concentration, e the electric charge, $\tau$ the relaxation time of carriers, and $m^*_{ab}$ and $m^*_c$ are the effective masses in the ab-plane and along the c-axis, respectively. The value of $\gamma$ thus obtained is 8.4 just above $T_c$, which is larger than $\gamma = 2.7$ estimated from the anisotropy of $H_{c2}$. This may be due to the large value of $\rho_c$ probably caused by the strong scattering of carriers by excess Fe. That is, $\tau$ in $\rho_c$ may be smaller than in $\rho_{ab}$.

## Ⅳ. Summary

Single crystals of FeSe$_{1-x}$Te$_x$ (0.5 ≤ x ≤1) were grown by the Bridgman method. The magnetic susceptibility measurements have revealed that single crystals of x = 0.5 – 0.9 annealed at 400℃ for 100 h in vacuum exhibit bulk superconductivity, though as-grown crystals of only x = 0.5 and 0.6 do. It seems to be due to the homogeneous distribution of Se and Te in a crystal through the annealing. The anisotropy of the electrical resistivity and $H_{c2}$ has been investigated for the annealed single-crystal of FeSe$_{1-x}$Te$_x$ with x = 0.6. The anisotropy in the electrical resistivity, $\rho_c/\rho_{ab}$, has been found to be 44 and 70 at 290 K and just above $T_c$, respectively. The anisotropy in $H_{c2}$ has been estimated from the angular dependence of $\rho_{ab}(H,\theta)$ under various constant magnetic fields just below $T_c$. The angular dependence of $H_{c2}$ thus obtained has been found to be explained in terms of the effective mass model, and the anisotropy parameter, $\gamma \equiv H_{c2}^{\parallel}/H_{c2}^{\perp}$, has been estimated as 2.7 just below $T_c$.

**Acknowledgments**

We would like to thank K. Takada and M. Ishikuro in Institute for Materials Research (IMR), Tohoku University, for their aid in the ICP-AES analysis. We also thank Y. Murakami in the Advanced Research Center of Metallic Glasses, IMR, Tohoku University, for his aid in the EPMA measurements. This work was supported by a Grant-in-Aid for Scientific Research from the Japan Society for the Promotion of Science.

**References**

1) Y. Kamihara, T. Watanabe, M. Hirano, and H. Hosono: J. Am. Chem. Soc. **130** (2008) 3296.

2) Z.-A. Ren, W. Lu, J. Yang, W. Yi, X.-L. Shen, Z.-C. Li, G.-C. Che, X.-L. Dong, L.-L. Sun, F. Zhou, and Z.-X. Zhao: Chin. Phys. Lett. **25** (2008) 2215.

3) F.-C. Hsu, J.-Y. Luo, K.-W. Yeh, T.-K. Chen, T.-W. Huang, P. M. Wu, Y.-C. Lee, Y.-L. Huang, Y.-Y. Chu, D.-C. Yan, and M.-K. Wu: Proc. Nati. Acad. Sci. **105** (2008) 14262.

4) S. Margadonna, Y. Takabayashi, Y. Ohishi, Y. Mizuguchi, Y. Takano, T. Kagayama, T. Nakagawa, M. Takata, and K. Prassides: Phys. Rev. B **80** (2009) 064506.

5) K.-W. Yeh, T.-W. Huang, Y.-L. Huang, T.-K. Chen, F.-C. Hsu, P. M. Wu, Y.-C. Lee, Y.-Y. Chu, C.-L. Chen, J.-Y. Luo, D.-C. Yan, and M.-K. Wu: Europhys. Lett. **84** (2008) 37002.

6) M. H. Fang, H. M. Pham, B. Qian, T. J. Liu, E. K. Vehstedt, Y. Liu, L. Spinu, and Z. Q. Mao: Phys. Rev. B **78** (2008) 224503.

7) S. Li, C. de la Cruz, Q. Huang, Y. Chen, J. W. Lynn, J. Hu, Y.-L. Huang, F.-C. Hsu, K.-W. Yeh, M.-K. Wu, and P. Dai: Phys. Rev. B **79** (2009) 054503.

8) W. Bao, Y. Qiu, Q. Huang, M. A. Green, P. Zajdel, M. R. Fitzsimmons, M. Zhernenkov, S. Chang, M. Fang, B. Qian, E. K. Vehstedt, J. Yang, H. M. Pham, L. Spinu, and Z. Q. Mao: Phys. Rev. Lett. **102** (2008) 247001.

9) S. B. Zhang, Y. P. Sun, X. D. Zhu, X. B. Zhu, B. S. Wang, G. Li, H. C. Lei, X. Luo, Z. R. Yang, W. H. Song, and J. M. Dai: Supercond. Sci. Technol. **22** (2009) 015020.

10) G. F. Chen, Z. G. Chen, J. Dong, W. Z. Hu, G. Li, X. D. Zhang, P. Zheng, J. L. Luo, and N. L. Wang: Phys. Rev. B **79** (2009) 140509(R).

11) B. C. Sales, A. S. Sefat, M. A. McGuire, R. Y. Jin, D. Mandrus, and Y. Mozharivskyj: Phys. Rev. B **79** (2009) 094521.

12) T. Taen, Y. Tsuchiya, Y. Nakajima, and T. Tamegai: Phys. Rev. B **80** (2009) 092502.

13) K. Kadowaki, A. A. Menovsky, and J. J. M. Franse: Physica B **165&166** (1990) 1159.

14) F. Lichtenberg, A. Catana, J. Mannhart, and D. G. Schlom: Appl. Phys. Lett. **60** (1992) 1138.

15) Y. Maeno, H. Hashimoto, K. Yoshida, S. Nishizaki, T. Fujita, J. G. Bednorz and F. Lichtenberg: Nature **372** (1994) 532.

16) R. C. Morris, R. V. Coleman, and R. Bhandari,: Phys. Rev. B **5** (1972) 895.

17) M. Fang. J. Yang, F. F. Balakirev, Y. Kohama, J. Singleton, B. Qian, Z. Q. Mao, H. Wang, and H. Q. Yuan: Phys. Rev. B **81** (2010) 020509(R).

18) N. Ni, S. L. Bud'ko, A. Kreyssig, S. Nandi, G. E. Rustan, A. I. Goldman, S. Gupta, J. D. Corbett,

A. Kracher, and P. C. Canfield: Phys. Rev. B **78** (2008) 014507.

19) H. Q. Yuan, J. Singleton, F. F. Balakirev, S. A. Baily, G. F. Chen, J. L. Luo, and N. L. Wang: Nature **457** (2009) 565.

20) Y. Iye, T. Tamegai, T. Sakakibara, T. Goto, N. Miura, H. Takeya, and H. Takei: Physica C **153-155** (1988) 26.

**Table I**. Compositions of FeSe$_{1-x}$Te$_x$ single crystals, and the superconducting transition temperature, T$_c$, and the magnetic transition temperature, T$_m$, for annealed single-crystals of FeSe$_{1-x}$Te$_x$. T$_c$ is defined as the onset temperature of the Meissner effect.

| x | nominal composition Fe : Se : Te | composition (ICP-AES) Fe : Se : Te | T$_c^{\text{annealed}}$ (K) | T$_m^{\text{annealed}}$ (K) |
|---|---|---|---|---|
| 0.5 | 1.03 : 0.50 : 0.50 | 1.03 : 0.46 : 0.54 | 14.0 | — |
| 0.6 | 1.03 : 0.40 : 0.60 | 1.00 : 0.39 : 0.61 | 14.2 | — |
| 0.7 | 1.03 : 0.30 : 0.70 | 1.03 : 0.28 : 0.72 | 14.3 | — |
| 0.8 | 1.03 : 0.20 : 0.80 | 1.03 : 0.19 : 0.81 | 13.8 | — |
| 0.9 | 1.03 : 0.10 : 0.90 | 1.02 : 0.08 : 0.92 | 11.3 | — |
| 1   | 1.13 : 0 : 1.00    | 1.09 : 0 : 1.00    | —    | 69 |

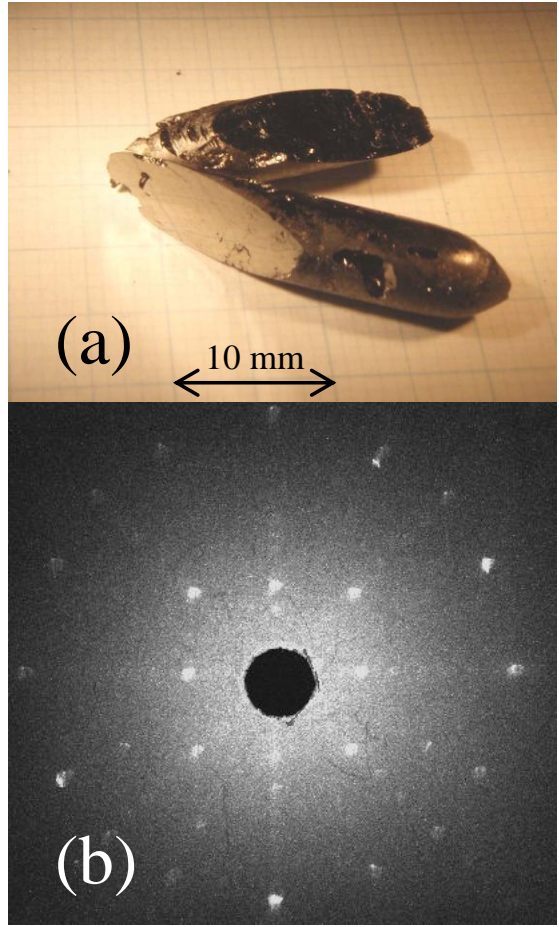

**Fig. 1**. (Color online) (a) Picture of as-grown single-crystals. (b) X-ray back-Laue photograph of an as-grown single-crystal of FeSe$_{1-x}$Te$_x$ with x = 0.8 in the x-ray perpendicular to the cleavage surface, namely, the ab-plane.

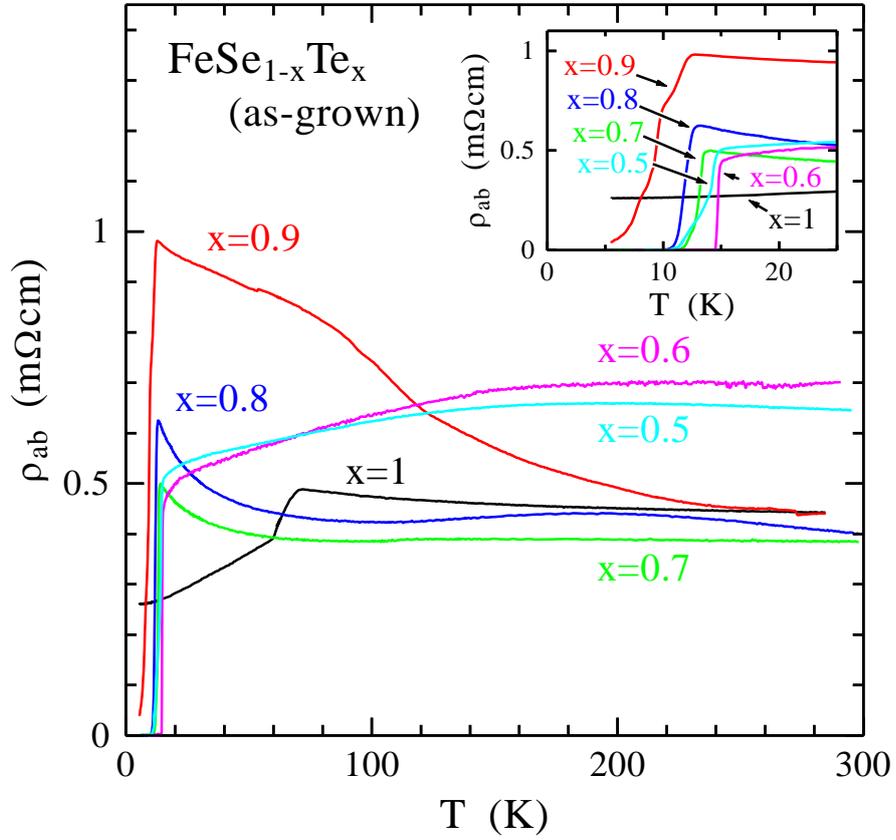

**Fig. 2.** (Color online) Temperature dependence of the in-plane resistivity, $\rho_{ab}$, of as-grown single-crystals of FeSe$_{1-x}$Te$_x$ ($0.5 \leq x \leq 1$). The inset is an enlarged plot of $\rho_{ab}$ at low temperatures.

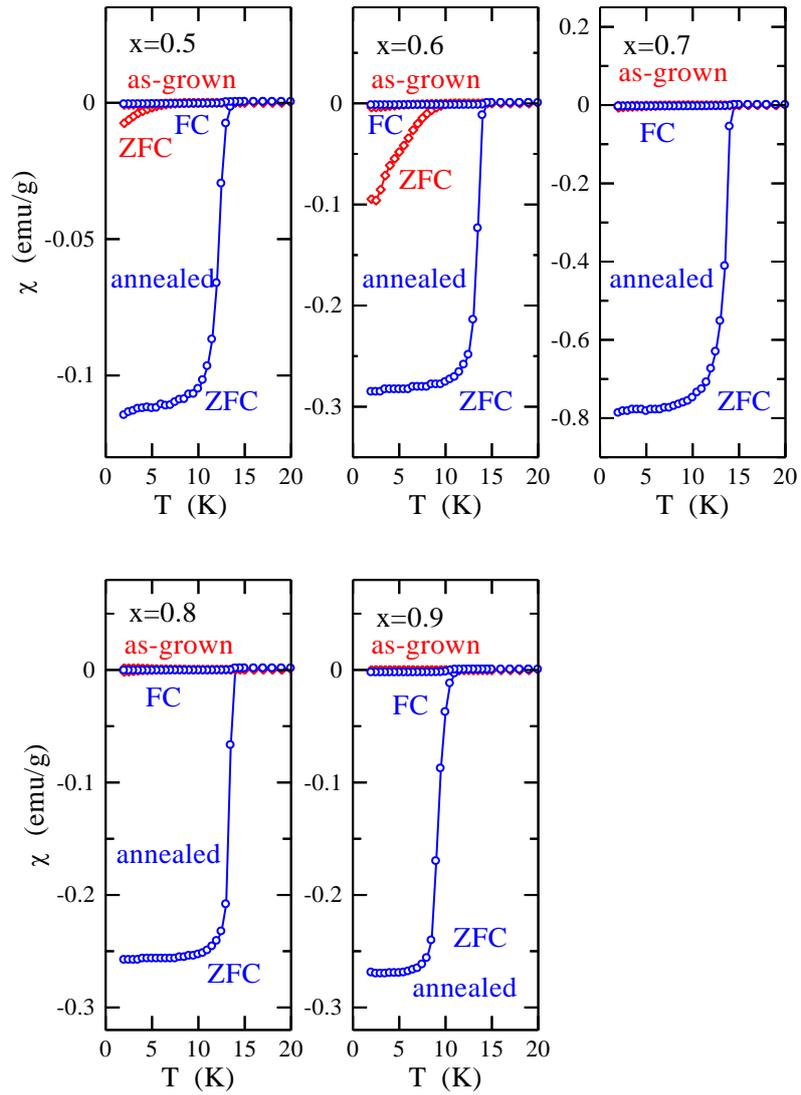

**Fig. 3**. (Color online) Temperature dependence of the magnetic susceptibility, χ, in a magnetic field of 1 mT parallel to the c-axis on zero-field cooling (ZFC) and field cooling (FC) for as-grown and annealed single-crystals of FeSe$_{1-x}$Te$_x$ with x = 0.5 - 0.9.

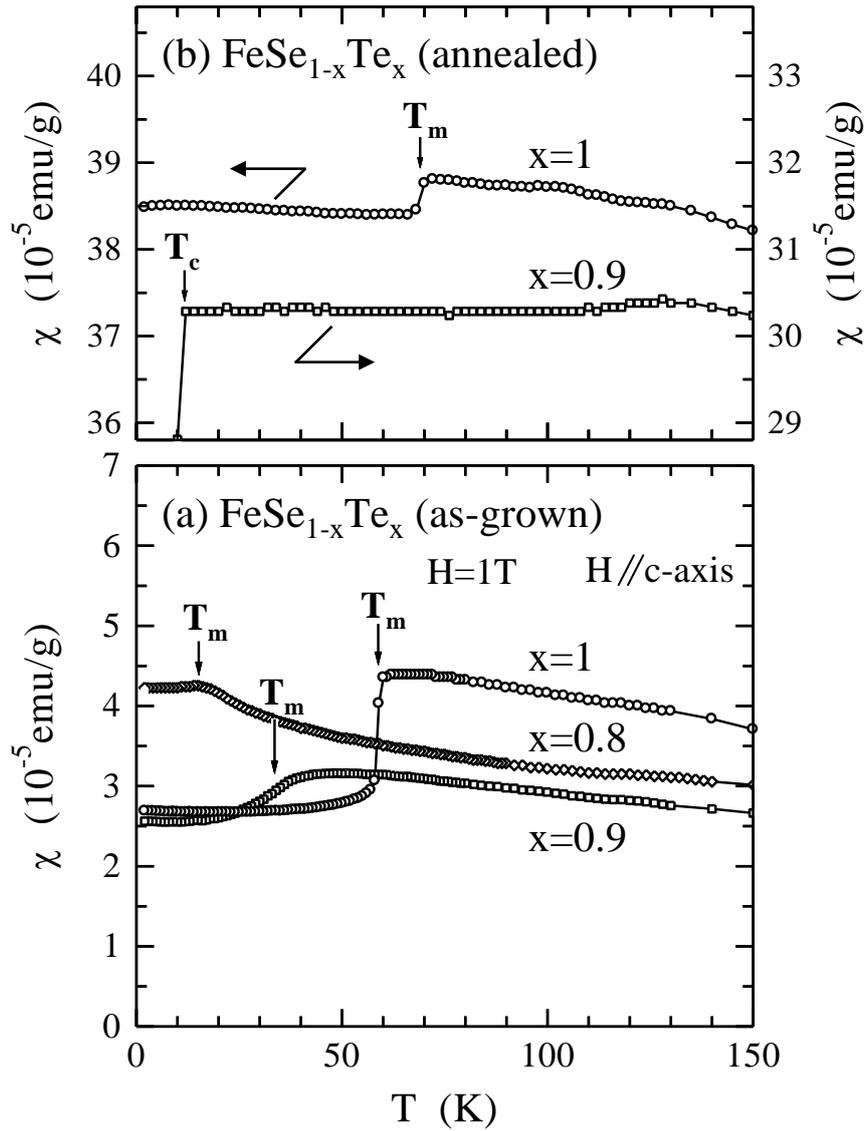

**Fig. 4**. Temperature dependence of the magnetic susceptibility, $\chi$, in a magnetic field of 1 T parallel to the c-axis for (a) as-grown and (b) annealed single-crystals of $FeSe_{1-x}Te_x$ with $x = 0.8 - 1$. Arrows indicate the magnetic transition temperature, $T_m$, and the superconducting transition temperature, $T_c$.

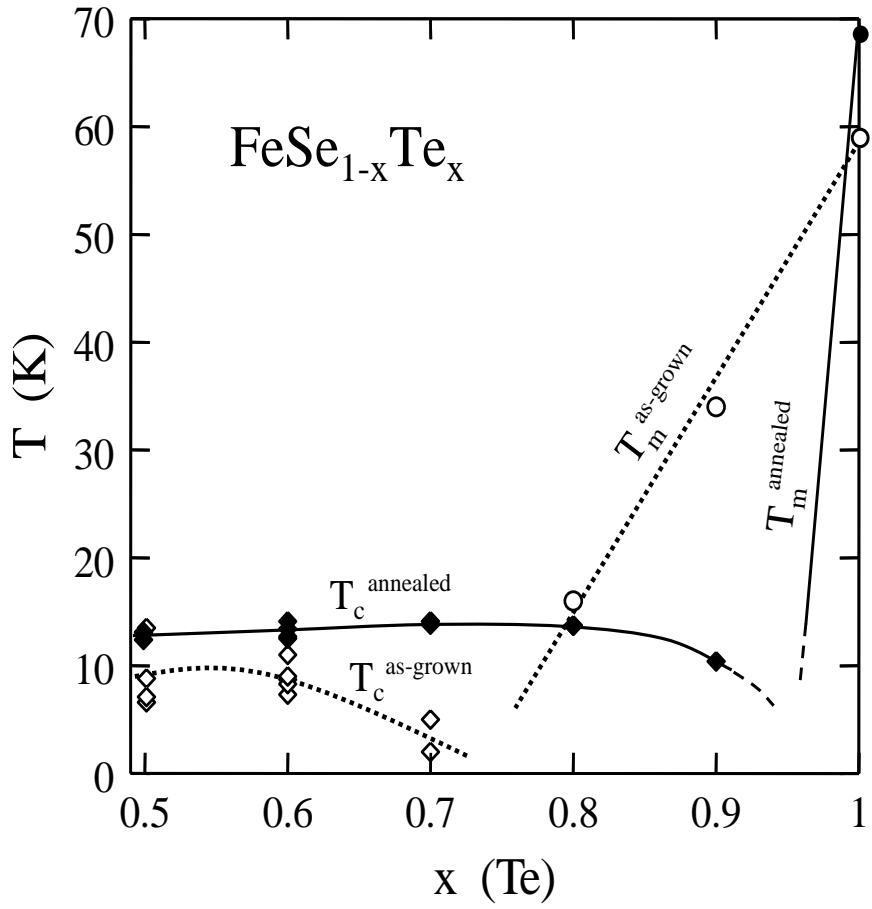

**Fig. 5**. Dependences on x of the superconducting transition temperature, $T_c$, and the magnetic transition temperature, $T_m$, for as-grown and annealed single-crystals of $FeSe_{1-x}Te_x$ ($0.5 \leq x \leq 1$).

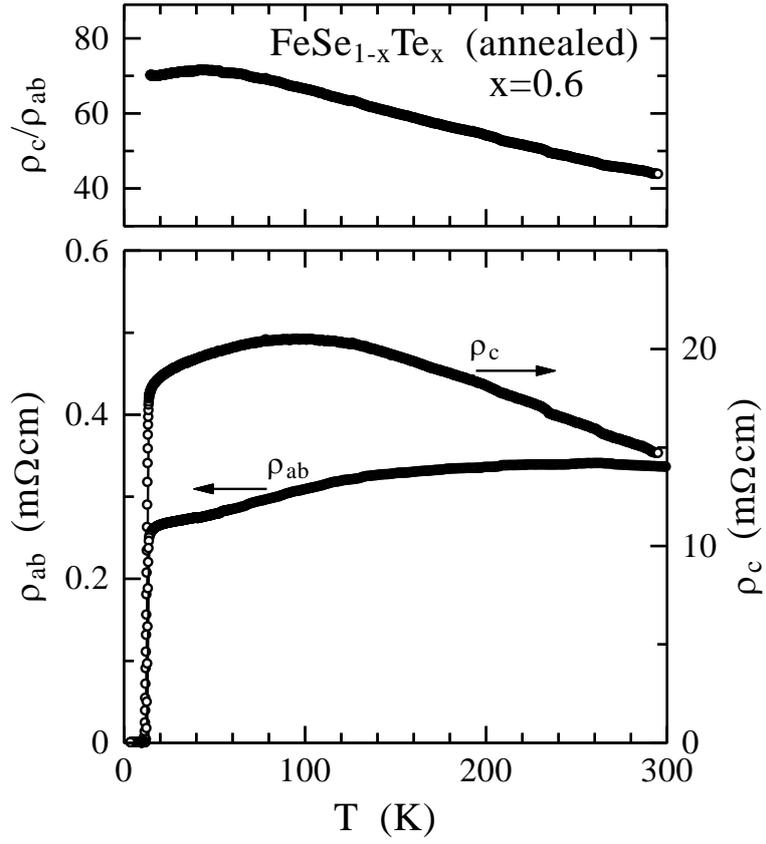

**Fig. 6**. Temperature dependence of the in-plane electrical resistivity, $\rho_{ab}$, and the out-of-plane one, $\rho_c$, (the bottom panel), and the resistivity ratio, $\rho_c/\rho_{ab}$, (the top panel) for the annealed single-crystal $FeSe_{1-x}Te_x$ with x = 0.6.

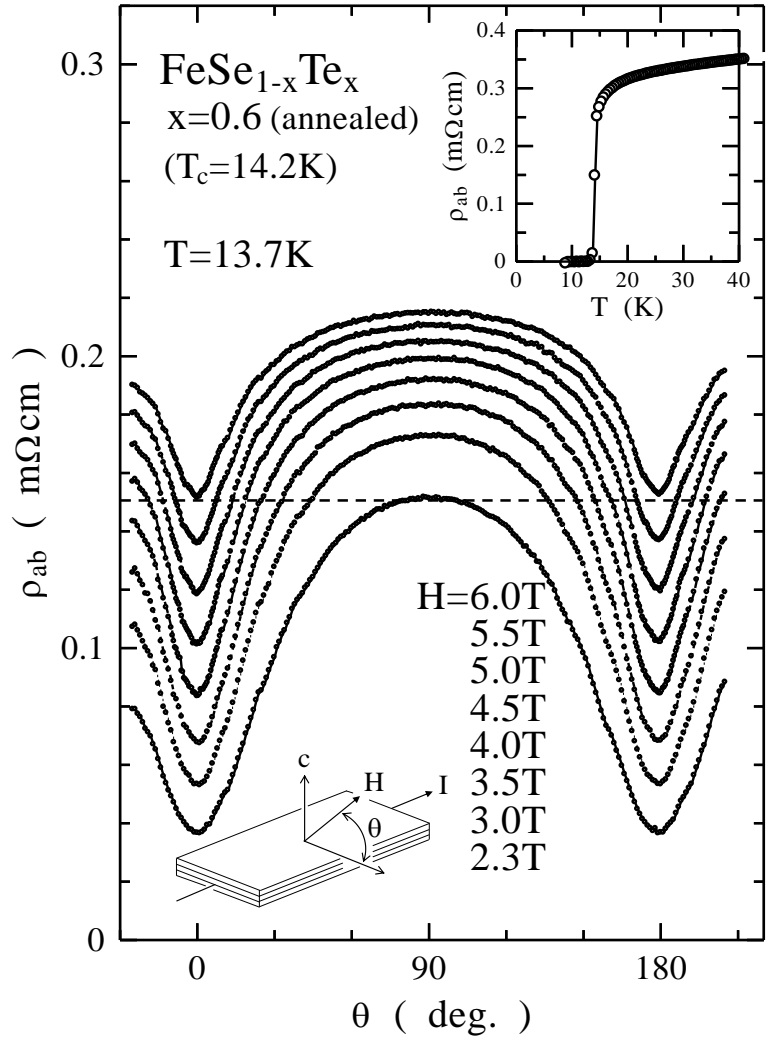

Fig. 7. Dependence of $\rho_{ab}$ on the angle between the ab-plane and the magnetic field direction, $\theta$, under various constant magnetic fields for the annealed single-crystal FeSe$_{1-x}$Te$_x$ with x = 0.6. The dashed straight line indicates the half of the normal-state value of $\rho_{ab}$ in zero field. The inset shows the resistive superconducting transition in $\rho_{ab}$ at low temperatures in zero field.

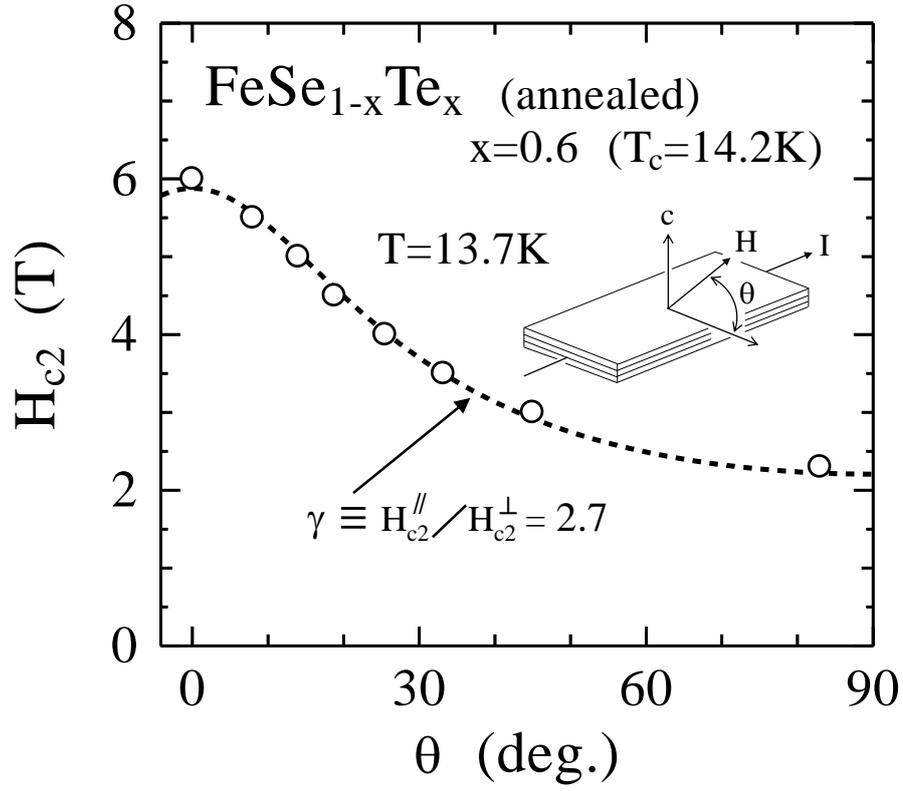

**Fig. 8**. Angular dependence of $H_{c2}$, defined at the half of the normal-state value as shown in Fig. 7, for the annealed single-crystal $FeSe_{1-x}Te_x$ with $x = 0.6$. The dashed curve indicates the best-fit result obtained using $H_{c2}(\theta) = H_{c2}^{//}(\cos^2\theta + \gamma^2\sin^2\theta)^{-1/2}$ with $H_{c2}^{//} = 5.9$ T and $\gamma = 2.7$.